\documentclass[prl,showpacs,twocolumn,floatfix,amsfonts]{revtex4}
\usepackage{graphicx,graphics,color,epsfig}
\usepackage{bm}
\usepackage{amsmath}
\usepackage{amssymb}

\newcommand{\bd}{{\bf d}}
\newcommand{\bB}{{\bf B}}

\newcommand{\bk}{{\bf k}}
\newcommand{\bv}{{\bf v}}

\newcommand{\br}{{\bf r}}
\newcommand{\bp}{{\bf p}}

\newcommand{\bA}{{\bf A}}
\newcommand{\ba}{{\bf a}}
\newcommand{\bJ}{{\bf J}}
\newcommand{\bE}{{\bf E}}

\newcommand{\Tr}{\,{\rm{Tr}\,}}
\newcommand{\tr}{\,{\rm{tr}\,}}
\newcommand{\bn}{{\bf \nabla}}

\newcommand{\beqa}{\begin{eqnarray}}
\newcommand{\eeqa}{\end{eqnarray}}

\renewcommand{\Re}{{\rm Re}}
\renewcommand{\Im}{{\rm Im}}
\begin{document}
\preprint{}
\title{Dipolar superfluidity in electron-hole bilayer systems}
\author{Alexander V. Balatsky,$^1$ Yogesh N. Joglekar,$^1$ and Peter B.
Littlewood$^{1,2}$} \affiliation{$^1$ Theoretical Division, Los
Alamos National Laboratory, Los
Alamos, New Mexico 87544, USA\\
$^2$ Cavendish Laboratory (TCM), University of Cambridge,
Cambridge CB3 0HE, United Kingdom}
\date{\today}
\begin{abstract}
Bilayer electron-hole systems, where the electrons and holes are created via 
doping and confined to 
separate layers, undergo excitonic condensation when distance between the 
layers is smaller than the typical distance between particles within a layer. 
We argue that the excitonic condensate is a novel dipolar superfluid in 
which the phase of the condensate couples to the {\it gradient} of the vector 
potential. We predict the existence of dipolar supercurrent which can be 
tuned by an in-plane magnetic field. Thus the dipolar superfluid offers an 
example of excitonic condensate in which the {\it composite} nature of its 
constituent excitons is manifest in the macroscopic superfluid state. We also 
discuss various properties of this superfluid including the role of vortices.
\end{abstract}
\maketitle

{\it Introduction:} Superfluidity was discovered by Kapitsa in 1938 as the 
ability of liquid Helium (He$^4$) to carry momentum without 
dissipation~\cite{kapitsa}. Its phenomenological theory was developed by 
Landau and collaborators, who introduced the notion of the condensate wave 
function as an order parameter that describes the superfluid component of the 
liquid~\cite{landau}. This superfluid component can carry momentum with no 
dissipation. The cornerstone of the Landau theory of superfluidity is the 
notion that phase of the superfluid condensate couples to the gauge potential 
and that the condensate current is given by $\bJ=\rho_s(\bn \phi-e\bA)$ where 
$\rho_s$ is a measure of the superfluid density, $\phi$ is the 
phase of the condensate wave function and $\bA$ is the gauge potential (we use
units such that $c=1=\hbar$). He$^4$, being neutral, does not carry 
electrical current although it can produce nontrivial gauge potentials by 
rotation. Another example of a neutral superfluid is an excitonic condensate. 
Electrons and holes in semiconductors can form (metastable) bound states 
called excitons which are expected to behave as neutral bosons at low 
densities, and therefore can undergo Bose-Einstein
condensation~\cite{moskalenko,keldysh,bilayer,rice,shev}. The condensate, 
being neutral, will posses properties similar to superfluidity. During the 
past few years electron-hole plasmas have been created by optically exciting 
the electrons from a valence band to the conduction band and then spatially 
confining the resulting electrons and holes to different quantum wells using
a static electric field~\cite{butov}. The investigation of their properties 
has been limited to photoluminescence measurements: processes which can probe 
phase-coherence but not superfluidity. In particular, they have not been 
investigated via transport measurements which can provide a {\it direct} 
signature of their superfluid properties.

It has been argued that quantum Hall {\it electron-electron} bilayers at total 
filling factor $\nu=1$ provide a realization of excitonic superfluid. These 
systems have been studied via transport; they show, among spectacular 
transport properties, a small but finite dissipation in the counterflow 
channel~\cite{em}. Here we focus on {\it electron-hole} systems in zero 
magnetic field perpendicular to the layers, discussing effects due to the 
presence of a physical dipole.

Recent developments in heterojunction fabrications open up an exciting 
possibility of electron-hole bilayer systems where the electrons reside in 
one layer and holes in the other layer separated by a distance $d$ 
($\sim$200\AA), and where the density of electrons or holes in individual 
layers can be adjusted using independent gates~\cite{sivan}. These 
systems have very weak but {\it nonzero} interlayer tunneling which allows 
the electrons and holes to couple with an in-plane magnetic field applied 
between the two layers. As a result of Coulomb attraction between electrons 
and holes, the excitonic condensate is expected to occur when the typical 
distance $r_s$ between electrons (holes) within a layer exceeds the distance 
$d$ between the two layers~\cite{loz,pbl,palo} (In contrast, bilayer quantum 
Hall systems can be mapped on to excitonic superfluids only near specific 
filling factors such as $\nu=1$). These systems offer an alternate view of 
the electron-hole excitonic condensate where the condensate is 
neutral, yet has a well defined {\it dipole moment} associated with each 
exciton. In this Letter, we argue that this excitonic condensate will 
represent a qualitatively new kind of superfluid where the condensate is 
neutral and carries no momentum density. We call this nominally neutral 
superfluid a {\it dipolar superfluid}. The dipole moment associated with each 
exciton in the condensate allows this liquid to couple to electromagnetic 
fields in a nontrivial fashion. We find that the phase of the dipolar 
superfluid couples to the {\it gradient of the gauge potential}. As a result, 
we predict that it 
will exhibit a neutral persistent dipolar current, consisting of equal and 
oppositely directed currents in the two layers, upon application of an in-plane
magnetic field $\bB_{||}$. Thus, the composite structure of excitons is 
manifest in the macroscopic superfluid state. In the 
following paragraphs, we present the hydrodynamics of such a superfluid 
based on the Ginzburg-Landau (GL) energy functional, state various predictions
which follow from it (including the existence of critical field 
$\bB^{c}_{||}$ above which the dipolar superfluidity is destroyed), and then 
briefly outline the derivation of the GL functional from a microscopic 
Hamiltonian. Because there is no long-range order in two dimensions, 
photoluminescence from a coherent exciton droplet is suppressed when the 
thermal length for phase fluctuations reaches the size of the 
droplet~\cite{pl}. On the other hand, the phase stiffness, which determines 
superfluidity, persists up to the Kosterlitz-Thouless (KT) transition 
temperature which is much higher. Therefore, a direct probe of phase 
stiffness can provide a better signature of excitonic condensation. Although 
the true nature of the superfluid phase transition (being KT-type) cannot be 
described by the standard mean-field theory, we will concentrate on 
temperatures much smaller than the KT-transition temperature, $T\ll T_{KT}$, 
where the Hartree-Fock mean-field description is applicable~\cite{vig}.

\begin{figure}[thbp]
\begin{center}
\vspace{1cm}
\includegraphics[width=2.8in]{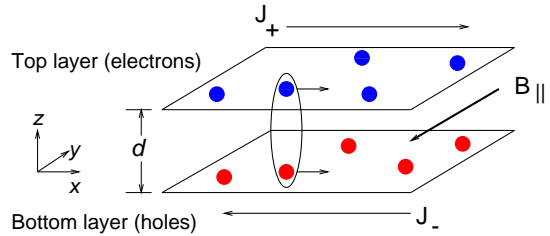}
\caption{Schematic bilayer electron-hole system. The electrons and holes 
form bound excitons which condense at low densities. We predict that an 
in-plane field $\bB_{||}$ will produce equal but opposite currents 
$\bJ_{+}$ and $\bJ_{-}$ in the two layers.} 
\label{fig: ehbilayer}
\end{center}
\vspace{-8mm}
\end{figure}

{\it GL Energy Functional and Effective Action:} Let us consider a bilayer 
system with electrons in the top layer and holes in the bottom layer. We 
introduce a notation where $\pm$ subscript corresponds to the top and bottom 
layer respectively. With the coordinate system shown in 
Fig.~\ref{fig: ehbilayer} we get $\br_\pm=\br\pm\bd/2$ where $\bd=d\hat{z}$ 
is a vector normal to the layers and $\br$ is a two-dimensional vector within 
the layer. We define the excitonic condensate order parameter as 
\beqa
\Delta(\br) =  \langle c^{\dag}_+(\br) c_-(\br)\rangle =
|\Delta(\br)| \exp\left[i\Phi(\br)\right] 
\label{eq: op} 
\eeqa
where $c^{\dag}_{\pm}(\br)$ creates an electron in the top (bottom) layer at 
position $\br$, and we have used the fact that $c_{-}(\br)=c^{\dag}_{h}(\br)$ 
where $c^{\dag}_{h}(\br)$ creates a {\it hole} at position $\br$ in the 
bottom layer. Upon a gauge transformation
$c_{\pm}(\br)\rightarrow\exp\left[ie\varphi_{\pm}(\br)\right]c_{\pm}(\br)$,
the order parameter $\Delta(\br)$ will transform as
$\Delta(\br)\rightarrow\exp\left[i\Phi(\br)\right]\Delta(\br)$
with 
\beqa 
\Phi(\br)\rightarrow\Phi(\br)-e\varphi_{+}(\br)+e\varphi_{-}(\br) 
\label{eq: gauge}
\eeqa 
where $-e<0$ is the electron charge. We will call the phase of the order 
parameter, $\Phi(\br)$, the {\it dipolar phase}. It is ``approximately'' 
charge-neutral since the condensate is an electron-hole condensate. However, 
the electron and the hole operators are always spatially separated: electrons 
are in the top layer and the holes are in the bottom layer. Therefore as one 
winds the phases of $c_{+}(\br)$ and $c_{-}(\br)$ in the same direction, 
since $\varphi_{\pm}(\br)=\varphi(\br\pm\bd/2)$, the phases that enter into 
the shift of the dipolar phase $\Phi(\br)$ {\it are not fully compensated}. 
This phase shift that enters in the gauge transformation of a nominally 
charge-neutral dipolar phase, Eq.(\ref{eq: gauge}), is crucial for the 
hydrodynamics of the dipolar superfluid. Now we determine the coupling of the 
dipolar phase $\Phi(\br)$ to external gauge potentials in the top and bottom 
layers, $\bA_{\pm}(\br)=\bA(\br\pm\bd/2)$. To be consistent with $U(1)$ gauge 
transformation 
$\bA_{\pm}(\br)\rightarrow\bA_{\pm}(\br)+\nabla\varphi_{\pm}(\br)$ the 
dipolar phase must transform as 
\beqa
\nabla\Phi(\br)\rightarrow \nabla\Phi(\br)-e\bA_+(\br)+e\bA_{-}(\br). 
\label{eq: gradphi} 
\eeqa 

The gauge potentials in top and bottom layers enter with opposite signs since 
they couple to oppositely charged electrons and holes respectively. 
Eq.(\ref{eq: gradphi}) shows that, in contrast to an ordinary neutral 
superfluid, {\it the phase of the dipolar superfluid couples to the 
difference of the gauge potentials between the two layers.}

The GL energy functional for the dipolar superfluid will depend on the order 
parameter $\Delta(\br)$, and for inhomogeneous state should depend only on 
the gauge-invariant combinations (\ref{eq: gradphi}) involving gradient of 
the dipolar phase $\Phi(\br)$. In particular, the gradient part of free 
energy will be given by
\beqa 
F=\frac{1}{2}\rho_{d}\int_{\br}\left[\nabla\Phi(\br)-e\bA_{+}(\br)+
e\bA_{-}(\br)\right]^2 
\label{eq: glenergy} 
\eeqa 
where $\rho_d$ is the dipolar superfluid density~\cite{cav}. It follows 
that the currents in the top and bottom layers are
\beqa 
\bJ_{\pm}(\br)=-\frac{\delta F}{\delta\bA_{\pm}(\br)}=\pm e\rho_d\left[
\nabla\Phi(\br)-e\ba(\br)\right]. 
\label{eq: sfcurrent} 
\eeqa 
Here we have introduced the antisymmetric and symmetric combinations of gauge
potentials, $\ba\equiv(\bA_{+}-\bA_{-})$ and 
${\cal A}\equiv(\bA_{+}+\bA_{-})$, which couple to the dipolar phase 
$\Phi(\br)$ and the total phase 
$\varphi_T=\varphi_{+}+\varphi_{-}$ respectively. Eq.(\ref{eq: sfcurrent}) 
implies that, as far as the dipolar condensate is concerned, there is no net 
electric current, $\bJ=(\bJ_{+}+\bJ_{-})=0$. This result is 
naturally expected since excitons do not carry any electric or mass current 
unless one breaks them apart. On the other hand, the excitonic condensate 
{\it does} carry a net dipolar current 
\beqa
\bJ_d(\br)=2e\rho_d\left[\nabla\Phi(\br)-e\ba(\br)\right].
\label{eq: dipolarcurrent} 
\eeqa 
Eq.(\ref{eq: dipolarcurrent}) is the central result of this paper. It has 
exactly the same form as a supercurrent in a superconductor,
$\bJ=e\rho_{s}(\nabla\phi-2e\bA)$. Therefore, in analogy with a 
superconductor, we expect persistent dipolar currents produced by the 
external gauge potential $\ba$. For a smoothly varying gauge potential, the 
antisymmetric combination 
$\ba(\br)=\bA(\br+\bd/2)-\bA(\br-\bd/2)\approx d\partial_{z}\bA(\br)$ can be 
tuned by varying a uniform in-plane magnetic field. To be specific, let us 
consider magnetic field $\bB_{||}=-B_{||}\hat{y}$ between the two layers, 
generated by gauge potential $\bA(\br,z)=(-B_{||}z,0,0)$. In this case,
assuming the dipolar condensate phase is uniform, we get 
\beqa
\bJ_{d} = 2{e^2}\rho_d dB_{||}\hat{x} 
\label{eq: pcurrent} 
\eeqa
Thus, we predict that {\it a uniform in-plane magnetic field will induce 
persistent and opposite currents in the top and the bottom layers in the 
direction perpendicular to magnetic field}. This is a direct consequence of 
the non-compensation of gauge potential acting on the electron and hole part 
of the dipolar condensate. 
Alternatively, one can view the dipolar persistent current $\bJ_d$ as arising 
from ``perfect diamagnetism'' of electrons and holes. Indeed, turning on the 
magnetic field $\bB_{||}$ produces in-plane electric fields $\bE_{\pm}$ which 
are equal and opposite. These electric fields accelerate electrons and holes 
{\it in the same direction} in respective layers. We emphasize that these 
time-dependent electric fields are physical and they cannot be gauged away in 
a simply-connected geometry. The condensate phase 
stiffness does not allow the resulting current to decay, thus giving rise to 
persistent dipolar supercurrent. Since the dipolar phase couples to 
$\ba\sim\bB_{||}$ it follows that the curl of the gauge potential 
$\nabla\times\ba$, {\it which will introduce vortices in the dipolar phase}, 
is determined by gradients in $\bB_{||}$ over the length-scale $d$ and is
necessarily small for externally applied fields. In addition, the constraint 
$\nabla\cdot\bB_{||}=0$ necessitates other gradients in the magnetic field to 
compensate for the gradients which induce vorticity in the gauge potential 
$\ba$. Thus creation of vortices in the dipolar phase requires nontrivial 
texture in the external magnetic field over very short length-scales. In this 
sense, the dipolar superfluid is robust against creation of vortices. This 
feature might make the particle-hole condensate more robust against 
vortex-induced dissipation in the counterflow channel.

It is straightforward obtain the effective action for dipolar superfluid from 
the GL energy functional, 
\beqa 
{\cal S}=\int_{\br,t}\left[n(\partial_t\Phi-ea_{0})-\frac{\rho_d}{2}
(\nabla\Phi-e\ba)^2-\frac{n^2}{2C}\right], 
\label{eq: action1} 
\eeqa 
where $a_0=(A_{0+}-A_{0-})$ is the difference between electrical potentials 
in the two layers, we have approximated the potential energy by a 
quadratic term with mass $1/C$ for the number fluctuations, and confined 
ourselves to a long-wavelength description~\cite{cav2}. The first 
term in the action is standard and results from the commutation relation 
between the dipolar condensate number and the phase, $[n,\Phi]=i$. 
Integrating out the massive fluctuations leads to the standard effective 
action for the phase, 
\beqa 
{\cal S}_\Phi=\int_{\br,t}\left[\frac{C}{2}(\partial_t \Phi)^2-
\frac{\rho_d}{2}(\bn\Phi)^2\right]. 
\label{eq: action2} 
\eeqa
Eq.(\ref{eq: action2}) implies that the phase fluctuation mode (superfluid 
sound mode) is a gapless collective mode with dispersion $\omega_{c}=v_ck$ 
where $v_c=\sqrt{\rho_d/C}$ is the collective mode velocity. It follows 
from the analogy with a neutral superfluid that a state with dipolar 
supercurrent $\bJ_d$ will be stable only if the superfluid velocity, 
defined by $\bJ_d=2en_d\bv_s=2e\bv_s/(\pi r_s^2)$, is smaller than the 
collective mode velocity $v_c$. Since the dipolar supercurrent is linearly 
proportional to $B_{||}$, we predict that for magnetic fields greater than 
a critical field 
$B^{c}_{||}=n_d/(ed\sqrt{\rho_d C})$, the dipolar superfluid state will be 
destroyed by collective phase fluctuations. For typical bilayer parameters 
($n_d\sim 10^{11}/$cm$^2$) we get the 
critical field $B^c_{||}\sim 100$T much larger than typical experimental 
field values. We can also define critical current as the dipolar current 
which leads to pair-breaking effects in which an exciton gives rise to an 
electron and a hole. This current is given by
$J^{c}_{d}=e\rho_d|\Delta|/v_F$ where $v_F$ is the Fermi velocity. Since the 
dipolar current is proportional to $B_{||}$, this criterion provides another 
bound on the maximum value of the in-plane field 
(for typical parameters we get $B_{||}^c\sim 10$T)~\cite{cav2}.

{\it Microscopic Theory:} We briefly describe the derivation of the dipolar 
current, Eq.(\ref{eq: dipolarcurrent}), from a microscopic model. The 
one-body Hamiltonian for an electron-hole bilayer system in the presence of 
gauge potentials is 
\beqa 
H_{1} & = & \sum_{\bk\bk'}\left[c^{\dag}_{+\bk'}\frac{1}{2m_e}
\left(\bp-e\bA_{+}\right)^2_{\bk'\bk}c_{+\bk}\right.\\ \nonumber &
& +\left.
c_{-\bk}\frac{1}{2m_h}\left(\bp+e\bA_{-}\right)^2_{\bk'\bk}
c^{\dag}_{-\bk}\right], 
\label{eq: micro} 
\eeqa 
where $m_e(m_h)$ is electron (hole) band-mass. The interaction term in the
microscopic Hamiltonian is a sum of the intralayer repulsive interaction, 
$V_A(\bk)=2\pi e^2/k$, and interlayer attractive interaction, 
$V_E(\bk)=-V_A(\bk)e^{-kd}$. We use Hubbard-Stratonovich transformation to 
introduce the order-parameter fields and obtain mean-field equations~\cite{no}.
The attractive interaction leads to the pairing of electrons and holes near 
the Fermi surface ($|\bk|\approx k_F$) and a nonzero dipolar condensate order 
parameter $\Delta=\langle c^{\dagger}_{+}c_{-}\rangle=\langle c^{\dagger}_{e}
c^{\dagger}_{h}\rangle$. In the absence of external gauge potentials, the 
mean-field Hamiltonian is given by 
\beqa
H_0=\sum_{\bk\sigma\sigma'}c^{\dagger}_{\bk\sigma}\left[\epsilon_{3\bk}
\tau^{3}+\Re\Delta_{\bk}\tau^{1}+\Im\Delta_{\bk}\tau^{2}\right]
_{\sigma\sigma'}c_{\bk\sigma'} 
\label{eq: mf} 
\eeqa
where $\sigma,\sigma'=\pm$ is the layer index, 
$\epsilon_{3}=(\epsilon_e+\epsilon_h)/2$ is the mean-band-energy and 
$\epsilon_{e(h)\bk}=\hbar^2k^2/2m_{e(h)}$ is the electron (hole) dispersion. 
In Eq.(\ref{eq: mf}) we have neglected a constant term proportional to
$\epsilon_0=(\epsilon_{e}-\epsilon_{h})/2$ since it vanishes for symmetric 
electron-hole bilayers, $m_e=m_h=m^{*}$, and does not qualitatively change 
the conclusions. The eigenvalues of the mean-field Hamiltonian are given by
$E_{\pm}(\bk)=\epsilon_{0\bk}\pm\sqrt{\epsilon_{3\bk}^2+|\Delta_{\bk}|^2}$
and the corresponding mean-field Green's function is
$G^{-1}_{0}(\bk,i\omega_n)=(i\omega_n-H_0)$. Now we will consider the 
effect of small gauge potential perturbations on the condensate state. To 
linear order, the change in the Hamiltonian is 
$\delta\epsilon_{e(h)}=\mp e\left[\bp\cdot\bA_\pm+\bA_\pm\cdot\bp\right] 
/(2m_{e(h)})$, which leads to the change in the Green's function, 
$G^{-1}=G^{-1}_0-\delta G^{-1}$, where 
\beqa 
\delta G^{-1}=-\frac{e}{2}\left\{\frac{1}{m_e}[\bp,\bA_{+}]_{+}-\frac{1}{m_h}
[\bp,\bA_{-}]_{+}\right\}. 
\label{eq: deltag1} 
\eeqa 
Note that the gauge potentials in the two layers have a relative minus sign
because the charges of the carriers in the two layers are opposite. This is 
also consistent with the GL energy functional description in which the 
dipolar phase is coupled to the 
{\it difference} of gauge potentials in the two layers. The change in the 
free energy due to a change in the gauge potential is given by 
$\delta F=-\Tr\left(G_0\delta G^{-1}G_0\delta G^{-1}\right)/2$ where ``Tr'' 
denotes the trace over space-time and layer-index degrees of 
freedom~\cite{no}. For symmetric electron-hole bilayers, in the 
long-wavelength limit, the matrix $\delta G^{-1}$ in layer-index-space is 
given by 
$\delta G^{-1}=-e\bk_F\cdot\left[\ba{\bf 1}+{\cal A}\tau^3\right]/(2m^{*})$. 
Evaluating the trace over space-time degrees leads to the dipolar superfluid 
current $\bJ_d$ 
\beqa 
\bJ_d=\frac{e^2v_F^2}{4}\tr \left[G_0\cdot{\bf 1}\cdot G_0
\left(\ba{\bf 1}+{\cal A}\tau^3\right) \right],
\label{eq: currentd}
\end{eqnarray}
with a similar expression for the total mass current $\bJ$. 
Note that only the term proportional $\ba{\bf 1}$ survives in the 
trace for the dipolar current $\bJ_d$. This reflects the fact that 
the dipolar condensate phase $\Phi$ responds to the antisymmetric 
combination of gauge potentials. Since we are expanding around $\bA=0$ 
and since the effect of $e\ba$ is the same as that of $\nabla\Phi$, only 
the gauge-invariant combination enters the expression for the dipolar 
current,
\beqa 
\bJ_d=\frac{2ev_F^2}{8}\tr[G_0{\bf 1}G_0{\bf1}]\left(\nabla\Phi-e\ba\right)
=2e\rho_d\left[\nabla\Phi-e\ba\right]. 
\label{eq: dpcurrent}
\eeqa 
Thus we recover Eq.(\ref{eq: dipolarcurrent}) with the appropriate definition 
of dipolar superfluid density from a microscopic calculation. In general, if 
we expand around a nonzero ${\cal A}$, which corresponds to nonuniform 
fields, we will also get a nonzero total mass current $\bJ$.

{\it Discussion:} The possibility of excitonic condensation in semiconductors 
has been discussed in the literature for a long time. Recent claims of 
observations of such condensates depend primarily on photoluminescence 
measurements and do not directly probe their superfluid properties. We have 
argued that excitonic condensates in electron-hole bilayer systems will 
present a qualitatively different superfluid in which the phase of the 
condensate is coupled to the {\it gradient} of the gauge potential. As a 
result we predict that such systems will develop persistent dipolar 
supercurrent with the application of an in-plane magnetic field. This 
supercurrent can be detected via separate contacts to the electron and the 
hole layers and will provide a {\it direct} signature of the superfluid 
properties of excitonic condensates. As a possible realization of this 
experiment in available samples, we propose the study of induced charges and 
voltages in each layer in response to an {\it ac} in-plane magnetic field 
with frequency $\omega$. In the absence of the excitonic condensate, Faraday 
induction will lead to a dipolar current $J_{d}\sim\omega B$ which in turn 
will induce equal and opposite charges $Q_{\pm}\sim\pm B$ in the two layers. 
In contrast, in the presence of condensate, the dipolar current will be 
$J_{d}\sim B$ and therefore the induced charges will be given by 
$Q_{\pm}\sim\pm B/\omega$, leading to a very different frequency dependence. 
In the current experimental setups, where the excitons are weakly confined in 
a parallel trap and recombine by optical emission, our analysis implies that 
the position of the spot, where recombinations take place, will oscillate 
with an applied oscillating in-plane magnetic field. 

We thank Nick Bonesteel, Jim Eisenstein, Mike Lilly, Allan MacDonald, and Kun 
Yang for useful discussions. This work was supported by LDRD at Los Alamos.

\end{document}